\documentclass[a4paper,11pt]{article}

\usepackage{amssymb,latexsym}

\begin{document}
\topmargin -10mm
\oddsidemargin 0mm

\renewcommand{\thefootnote}{\fnsymbol{footnote}}
\newcommand{\nn}{\nonumber\\}
\begin{titlepage}
\begin{flushright}
hep-th/0602156
\end{flushright}
\vspace{15mm}
\begin{center}
{\Large \bf  Friedmann Equations of FRW Universe in Scalar-tensor
Gravity, $f(R)$ Gravity and First Law of Thermodynamics }
\vspace{22mm}

{\large
  M. Akbar\footnote{Email address: akbar@itp.ac.cn} and
  Rong-Gen Cai~\footnote{Email address: cairg@itp.ac.cn }}\\
\vspace{8mm} { \em Institute of Theoretical Physics, Chinese
Academy of Sciences,\\
 P.O. Box 2735, Beijing 100080, China}
\vspace{10mm}

\end{center}
\vspace{10mm} \centerline{{\bf{Abstract}}}
 \vspace{15mm}
In the paper, hep-th/0501055 (R.G. Cai and S.P. Kim, JHEP {\bf
0502}, 050 (2005)), it is shown that by applying the first law of
thermodynamics to the apparent horizon of an FRW universe and
assuming the geometric entropy given by a quarter of the apparent
horizon area, one can derive the Friedmann equations describing
the dynamics of the universe with any spatial curvature; using the
entropy formula for the static spherically symmetric black holes
in Gauss-Bonnet gravity and in more general Lovelock gravity,
where the entropy is not proportional to the horizon area, one can
also obtain the corresponding Friedmann equations in each gravity.
In this note we extend the study of hep-th/0501055 to the cases of
scalar-tensor gravity and $f(R)$ gravity, and discuss the
implication of results.

\end{titlepage}

\newpage
\renewcommand{\thefootnote}{\arabic{footnote}}
\setcounter{footnote}{0} \setcounter{page}{2}

\vspace{.5cm}
 Since the discovery of black hole thermodynamics in 1970s, physicists
 have
been speculating that there should be some relationship between
 thermodynamics and Einstein equations because the
horizon area (geometric quantity) of black hole is associated with
its entropy (thermodynamical quantity),  the surface gravity
(geometric quantity) is related with its temperature
(thermodynamical quantity) in black hole
thermodynamics~\cite{BekHaw}, and these quantities satisfy the
first law of thermodynamics. In 1995, Jacobson \cite{A1} was
indeed able to derive Einstein equations by applying the first law
of thermodynamics $\delta Q = TdS$ together with proportionality
of entropy to the horizon area of the black hole. He assumed that
this relation holds for all Rindler causal horizons through each
space time point with $\delta Q$ and $T$ interpreted as the energy
flux and Unruh temperature seen by an accelerated observer just
inside the horizon. On the other hand,
 Verlinde \cite{Ver} found that for a radiation
dominated Friedmann-Robertson-Walker (FRW) universe, the Friedmann
equation can be rewritten in the same form as the Cardy-Verlinde
formula, the latter is an entropy formula for a conformal field
theory in a higher dimensional space-time. Note that the radiation
can be described by a conformal field theory. Therefore, the
entropy formula describing the thermodynamics of radiation in the
universe has the same form as that of the Friedmann equation,
which describes the dynamics of spacetime. In particular, when the
so-called Hubble entropy bound is saturated, these two equations
coincide with each other (for a more or less complete list of
references on this topic see, for example, \cite{cm}). Therefore,
Verlinde's observation further indicates some relation between
thermodynamics and Einstein equations. For related discussions on
the relation between thermodynamics and Einstein equations see
also \cite{pad}.

 Frolov and Kofman in \cite{FK} employed the
approach proposed by Jacobson~\cite{A1} to a quasi-de Sitter
geometry of inflationary universe, where they calculated the
energy flux of a background slow-roll scalar field (inflaton)
through the quasi-de Sitter apparent horizon and used the first
law of thermodynamics
\begin{equation}
\label{first}
 -dE=TdS,
\end{equation}
where $dE$ is the amount of the energy flow through the apparent
horizon. Although the topology of the local Rindler horizon in
Ref. \cite{A1} is quite different from that of the quasi-de Sitter
apparent horizon considered in Ref. \cite{FK}, it was found that
this thermodynamic relation reproduces one of the Friedmann
equations with the slow-roll scalar field. It is assumed in their
derivation that
\begin{equation}
\label{1eq3}
 T=\frac{H}{2\pi}, \ \ \ S= \frac{\pi}{GH^2},
\end{equation}
where $H$ is a slowly varying Hubble parameter. Danielsson in
\cite{Dan} obtained the Friedmann equations, by applying the
relation $\delta Q=TdS$ to a cosmological horizon and by
calculating the heat flow through the horizon in an expanding
universe in an acceleration phase and assuming the same form for
the temperature and entropy of the cosmological horizon as those
in Eq. (\ref{1eq3}). Furthermore, Bousso \cite{Bousso} recently
considered thermodynamics in the Q-space (quintessence dominated
spacetime). Because the equation of state of quintessence is in
the range of $-1 < \omega <-1/3$, the universe undergoes an
accelerated expansion, and thus the cosmological event horizon
exists in  the Q-space. However, Bousso argued that a
thermodynamic  description of the horizon is approximately valid
and thus it   would not matter much whether one uses the apparent
horizon or the event horizon. Indeed, for the Q-space, the
apparent horizon  radius $R_A$ differs from the event horizon
radius $R_E$ only by a small quantity: $R_A/R_E=1-\epsilon$, where
$\epsilon = 3(\omega +1)/2$. Using the relations
  \begin{equation}
  \label{1eq4}
  T = \frac{1}{2\pi R_A}, \ \ \  S = \frac{\pi R^2_A}{G},
  \end{equation}
Bousso showed that  the first law (\ref{first}) of thermodynamics
holds at the apparent horizon of the Q-space.  While these authors
\cite{FK,Dan,Bousso} dealt with different aspects of the relation
between the first law of thermodynamics and Friedmann equations,
they considered only a {\it flat} FRW universe. On the other hand,
in modern cosmology, there are several different concepts of
horizons: particle horizon, Hubble horizon, event horizon and
apparent horizon. In a previous paper one of present authors and
Kim~\cite{A5} clarified some subtleties in deriving the Friedmann
equations from the first law of thermodynamics and have argued
that this exactly holds only for apparent horizon in the FRW
universe. They were able to derive the Friedmann equations of an
(n+1)-dimensional FRW universe with any spatial curvature by
applying the first law of thermodynamics to the apparent horizon.
Also by employing the entropy formula of static spherically
symmetric black holes in Gauss-Bonnet gravity and in more general
Lovelock gravity, they obtained the corresponding Friedmann
equations in each gravity. Note that in these higher order
derivative gravitational theories, the entropy of black holes does
not obey the area formula. For the related discussion see
\cite{Cal,Wang}.

 So it is obviously of great interest to see whether the above method is valid
 in general, or to what extent one can derive the Friedmann
 equations in a certain gravitational theory, given the relation
 between the entropy and horizon area of black holes in that
 theory. For this purpose, in this note we generalize the work
 in hep-th/0501055 to the scalar-tensor gravity and f(R)
 gravity and want to know whether the method developed
  in \cite{FK,Dan,Bousso,A5} holds or not in these two gravitational
  theories. Note that on the one hand, in these two theories black hole entropy
  does not obey the area formula of horizon; one the other hand,
  there are new degrees of freedom in these two theories (In the
  scalar-tensor gravity, there is an additional scalar field mediating
  the gravitational interaction; for the $f(R)$ gravity, there are more than
  second-order derivatives with respect to metric in the equations of motion).

  Let us start with the
$(n+1)$-dimensional FRW universe with metric
\begin{equation}\label{2}
ds^{2} = -dt^2 + a^2(t)\gamma_{ij}~ dx^{i} dx^{j},
\end{equation}
where the $n$-dimensional spatial hypersurface with negative, zero
or positive curvature is parameterized by $k = -1, 0$ and $ 1$,
respectively and the metric $\gamma_{ij}$ are given by
\begin{equation}\label{3}
\gamma_{ij} = \frac{dr^{2}}{1-k r^{2}} + r^{2}d\Omega^{2}_{n-1}.
\end{equation}
Here $d\Omega^{2}_{n-1}$ is the metric of an $(n-1)$-dimensional
sphere with unit radius. If $R_{ij}$ is the Ricci tensor given by
the metric $\gamma_{ij}$, one has $R_{ij}$ = $(n-1)k \gamma_{ij}$.
The Lagrangian of the scalar-tensor gravity can be written as

\begin{equation}\label{4}
L = \frac{f(\phi) R}{16 \pi G} - \frac{1}{2} g^{\mu\nu}
\partial_{\mu}\phi
\partial_{\nu}\phi - V(\phi) + L_{m},
\end{equation}
where $R$ is the Ricci curvature scalar of the space time, $\phi$
is the scalar field , $f(\phi)$ is an arbitrary function of the
scalar field $\phi$, and $L_{m}$ represents the Lagrangian for
matter fields in the universe. Varying the action Eq.~(\ref{4})
with respect to the two dynamical variables $g_{\mu\nu}$ and
$\phi$ yields the field equations
\begin{equation}\label{5}
G_{\mu\nu} = \frac{8 \pi G}{f(\phi)} \left(\partial_{\mu}\phi
\partial_{\nu} \phi - \frac{1}{2} g_{\mu\nu} (\partial \phi)^{2} -
g_{\mu\nu} V(\phi) -g_{\mu\nu} \nabla^{2} f + \nabla_{\mu}
\nabla_{\nu} f + T^{m} _{\mu\nu}\right ),
\end{equation}
and
\begin{equation}\label{6}
\nabla^{2}\phi - V^{\prime}(\phi) + \frac{1}{2} f'(\phi) R = 0,
\end{equation}
where $G_{\mu\nu}$ is the Einstein tensor and $T^{m}_{\mu\nu}$ is
the energy-momentum tensor of the matter fields. Assume that
$T^{m}_{\mu\nu}$ has the form of the energy-momentum tensor of a
perfect fluid
\begin{equation}\label{7}
T^{m}_{\mu\nu} = (\rho + p) U_{\mu} U_{\nu} + pg_{\mu\nu}
\end{equation}
where $U^{\mu}$ is the four velocity of the fluid, $\rho$ and $p$
are the energy density and the pressure of the fluid,
respectively. Note that the equation (\ref{5}) can also be written
as
\begin{equation}\label{8}
G_{\mu\nu} = 8 \pi G \tilde{T}_{\mu\nu} = \frac{8 \pi G}{f(\phi)}
( T^{m}_{\mu\nu} + T^{\phi}_{\mu\nu}).
\end{equation}
Solving equation (\ref{8}) for time-time and space-space
components one gets the Friedmann equations
\begin{equation}\label{9}
H^{2} + \frac{k}{a^{2}}= \frac{16 \pi G}{n(n-1)f} \left(
\frac{\dot{\phi^{2}}}{2} + V(\phi) - nH\dot{f} + \rho \right ),
\end{equation}
\begin{equation}\label{10}
\dot{H} - \frac{k}{a^{2}} = -\frac{8 \pi G}{(n-1)f}\left
(\dot{\phi}^{2} + \ddot{f} - H \dot{f} + \rho + p \right ),
\end{equation}
where the dots  represent derivatives with respect to the cosmic
time $t$, $H$ denotes the Hubble parameter, $H = \dot{a} / a$.

The metric (\ref{2}) can be written as
\begin{equation}\label{11}
ds^{2} = h_{ab}dx^{a}dx^{b} + \tilde{r}^{2}d\Omega^{2}_{n-1}
\end{equation}
where $\tilde{r}= a(t)r$ and $x^{0}=t$, $x^{1}=r$ and the two
dimensional metric $h_{ab} = {\rm diag}(-1, 1/(1-kr^{2}))$. The
dynamical apparent horizon is determined by the relation
$h^{ab}\partial_{a}\tilde{r}\partial_{b}\tilde{r}=0$, which yields
\cite{A6}
\begin{equation}\label{12}
\tilde{r}_{A} = 1/\sqrt{H^{2}+k/a^{2}}.
\end{equation}
The apparent horizon has been argued to be a causal horizon for a
dynamical space time and is associated with gravitational entropy
and surface gravity~\cite{A7}. Thus for our purpose, following
\cite{A5}, we here apply the first law of thermodynamics to the
apparent horizon. We define the work density $W$~\cite{A7}.
\begin{equation}\label{13}
W = -(1/2) T^{ab}h_{ab},
\end{equation}
and the energy supply vector
\begin{equation}\label{14}
\psi_{a} = T^{b}_{a}\partial_{b}\tilde{r} +
W\partial_{a}\tilde{r},
\end{equation}
where $T_{ab}$ is the projection of the $(n+1)$-dimensional
energy-momentum tensor $T_{\mu\nu}$ in FRW universe in the normal
direction of the $(n-1)$-dimensional sphere. The work density at
the
 apparent horizon should be regarded as the work done by
a change of the apparent horizon, while the energy-supply on the
horizon is the total energy flow through the apparent horizon. Now
according to the thermodynamics the entropy is associated with
heat flow as $\delta Q =TdS$ and the heat flow to the system is
related to the change of the energy of the system. Thus the
entropy is finally associated with the energy supply term and the
heat flow $\delta Q$ through the apparent horizon during
infinitesimal time interval, $dt$, is the amount of energy
crossing the apparent horizon during that time interval, i.e.
$\delta Q = -dE$, is the change of the energy inside the apparent
horizon. For the perfect fluid we find the energy supply vector
\begin{equation}\label{15}
\psi_{a}=\left (-\frac{1}{2}(\rho + p)H \tilde r,\
\frac{1}{2}(\rho + p)a\right ).
\end{equation}
Then the amount of energy crossing the apparent horizon during
$dt$, is given by
\begin{equation}\label{16}
-dE = -A\psi = A(\rho + p) H\tilde{r}_{A} dt
\end{equation}
where $A = n\Omega_{n}\tilde{r}_{A}^{n-1}$ is the area of the
apparent horizon and $\Omega_{n} = \pi^{n/2} / \Gamma(n/2 + 1)$.
In this scalar-tensor gravity theory, the entropy $S$ of black
holes has the form~\cite{CM1}
\begin{equation}\label{17}
S = \left. \frac{A f(\phi)}{4G} \right |_{\rm horizon},
\end{equation}
where $A$ is the horizon area of black holes in the scalar-tensor
theory. Starting from the entropy (\ref{17}) and assuming that the
apparent horizon has the temperature as in \cite{A5}
 \begin{equation} \label{temp}
 T= \frac{1}{2\pi \tilde r_A},
\end{equation}
we are no able to derive the Friedmann equations for the
scalar-tensor gravity (\ref{2}).
 However if we still take ansatz, $T = 1 / 2\pi\tilde{r}_{A}$ and $S = A / 4G$,
regard $\tilde{T}_{\mu\nu} = (T^{m}_{\mu\nu} + T^{\phi}_{\mu\nu})
/ {f(\phi)}$ as the effective energy-momentum tensor of the
 matters in the universe, and apply the first law of
thermodynamics, $-dE = T dS$, to the apparent horizon of the FRW
universe in the scalar-field gravity, we can obtain
\begin{equation}\label{18}
\dot{H} - \frac{k}{a^{2}} = -\frac{8 \pi G}{(n-1)f}\left
(\dot{\phi}^{2} + \ddot{f} - H \dot{f} + \rho + p \right ).
\end{equation}
This is nothing but one of the Friedmann equations describing an
(n+1)-dimensional FRW universe with any spatial curvature in the
scalar-tensor gravity. The continuity equation for the effective
perfect fluid in the scalar-tensor gravity is
\begin{equation}\label{19}
\dot{\tilde{\rho}} + n H (\tilde{\rho} + \tilde{p}) = 0,
\end{equation}
where we have defined the energy density $\tilde{\rho}$ and
pressure $\tilde{p}$ in the scalar-tensor gravity
\begin{equation}\label{20}
\tilde{\rho} = (\frac{\dot{\phi}^{2}}{2} + V -nH\dot{f} + \rho)/f,
\end{equation}
and
\begin{equation}\label{21}
\tilde{p} = (\frac{\dot{\phi}^{2}}{2} - V + (n-1)H\dot{f} +
\ddot{f} + p)/f,
\end{equation}
respectively. Using Eqs. (\ref{19}), (\ref{20}) and (\ref{21})
into Eq.~(\ref{18}) and integrating the resulting equation we
finally get
\begin{equation}\label{22}
H^{2} + \frac{k}{a^{2}}= \frac{16 \pi G}{n(n-1)f}
\left(\frac{\dot{\phi^{2}}}{2} + V(\phi) - nH\dot{f} + \rho
\right).
\end{equation}
This is another Friedmann equations (\ref{9}) in the scalar-tensor
gravity. Note that here we have set the integration constant be
zero. Thus we have derived the Friedmann equations of FRW universe
in the scalar-tensor gravity.

 Now we turn to the so-called $f(R)$ theory. Here $f(R)$ is a
 function of the curvature scalar $R$.
  The $f(R)$ gravity is a non-linear
gravitational theory or higher order derivative gravitational
theory since its equations of motion contain more than
second-order derivatives of metric. The Hilbert-Einstein action of
an $(n+1)$-dimensional $f(R)$ gravity in the presence of matters
can be written as
\begin{equation}\label{23}
S = \int d^{n+1}x \sqrt{-g}\left (\frac{f(R)}{16 \pi G} + L_{m}
\right ).
\end{equation}
The variational principle gives equations of motion
\begin{equation}\label{24}
G_{\mu\nu} = 8 \pi G \tilde{T}_{\mu\nu} = 8 \pi G \left (
\frac{1}{f^{\prime}} T^{m}_{\mu\nu} + \frac{1}{8 \pi G}
T^{curv}_{\mu\nu} \right ),
\end{equation}
where $T^{m}_{\mu\nu}$ and $T^{curv}_{\mu\nu}$ are given by
\begin{equation}\label{25}
T^{m}_{\mu\nu} = (\rho + p) U_{\mu} U_{\nu} + p g_{\mu\nu},
\end{equation}
and
\begin{equation}\label{26}
T^{curv}_{\mu\nu} = \frac{1}{f^{\prime}(R)} \left (g_{\mu\nu}
(f(R) - R f^{\prime}(R)) / 2 + \nabla_{\mu} \nabla_{\nu}
f^{\prime}(R) -g_{\mu\nu} \nabla^{2} f^{\prime}(R) \right ),
\end{equation}
respectively.  We define $T_{\mu\nu}^{total}$ as
\begin{equation}\label{27}
T^{total}_{\mu\nu} = \frac{1}{f^{\prime}} T^{m}_{\mu\nu} +
\frac{1}{8 \pi G} T^{curv}_{\mu\nu},
\end{equation}
and
\begin{equation}\label{28}
\rho_{total} = \frac{1}{f^{\prime}} \rho_{m} + \frac{1}{8 \pi G}
\rho_{curv},
\end{equation}
\begin{equation}\label{29}
p_{total} = \frac{1}{f^{\prime}} p_{m} + \frac{1}{8 \pi G}
p_{curv}.
\end{equation}
Here $\rho_{curv}$ and $p_{curv}$ are given by
\begin{equation}\label{30}
\rho_{curv} = \frac{1}{f^{\prime}} (-(f - R f^{\prime}) / 2 - nH
f^{\prime\prime} \dot{R}),
\end{equation}
\begin{equation}\label{31}
p_{curv} = \frac{1}{f^{\prime}} ( (f - R f^{\prime}) / 2 +
f^{\prime\prime} \ddot{R} + f^{\prime\prime\prime} \dot{R}^{2} +
n(n-1) f^{\prime\prime} \dot{R}),
\end{equation}
in the FRW universe.  Solving equations (\ref{24}) in the FRW
universe (\ref{2}), we get
\begin{equation}\label{32}
H^{2} + \frac{k}{a^{2}} = \frac{16 \pi G}{n(n-1)}
 \left (\frac{\rho_{m}}{f^{\prime}} + \frac{\rho_{curv}}{8 \pi G} \right
),
\end{equation}
and
\begin{equation}\label{33}
\dot{H} - \frac{k}{a^{2}} = \frac{-8 \pi G}{n-1}\left
(\rho_{total} + p_{total} \right ).
\end{equation}
These are two Friedmann equations for the $f(R)$
gravity~\cite{Cap}.

 In this gravity theory the black hole entropy has a relation to its horizon
 area~\cite{Wald}
\begin{equation}\label{34}
S =\left. \frac{A f^{\prime}(R)}{4 G} \right |_{\rm horizon},
\end{equation}
where $A$ is the horizon area of black holes in the $f(R)$ theory.
Using the entropy formula (\ref{34}) and the ansatz for the
temperature (\ref{temp}) for the apparent horizon, and applying
the first law (\ref{first}) to the apparent horizon, we fail to
re-produce corresponding Friedmann equations  in the $f(R)$
gravity. However, if we take $\tilde{T}_{\mu\nu}$ as the effective
energy-momentum tensor of matters in the $f(R)$ gravity and still
use the ansatz $S = A / 4 G$ and $T = 1 /2 \pi r_{A}$, we  can
successfully  derive the corresponding Friedmann equations (Eqs.
(\ref{32}) and (\ref{33})) for the $f(R)$ gravity.

As conclusion, in hep-th/0501055, assuming the area formula of
black hole entropy in Einstein gravity and taking the ansatz for
the temperature (\ref{temp}) for apparent horizon, it was shown
that the Friedmann equations of the FRW universe can be derived by
applying the first law of thermodynamics to the apparent horizon
for an FRW universe with any spatial curvature. Employing the
entropy relation of black holes to horizon area in Gauss-Bonnet
gravity and in the more general Lovelock gravity, one also can get
the corresponding Friedmann equations in these two gravities.
These results combining with those in \cite{FK,Dan,Bousso}
manifestly demonstrate the gravitational holographic properties:
Einstein equations of gravity imply that the entropy inside a
causal horizon is proportional to  its horizon area; on the other
hand, if assuming a relation between the entropy and horizon area,
one can obtain corresponding equations of motion of gravitational
field. By using the first law of thermodynamics, whether can we
always obtain corresponding Friedmann equations in any
gravitational theory, given the geometric entropy relation to the
horizon in that gravitational theory?  To see this, in this note
we have shown that it does not naively hold: if we take the
entropy relation of black holes in the scalar-tensor theory and
$f(R)$ theory, applying it to the apparent horizon, we are not
able to re-produce corresponding Friedmann equations in these two
theories. However, if we still take the area formula of geometric
entropy and regards some terms concerning the scalar in the
scalar-tensor theory and extra curvature terms in the $f(R)$
theory as effective matters, we have indeed obtained corresponding
Friedmann equations in the scalar-tensor gravity and $f(R)$
gravity. The reason of failure seems that due to the new
additional degrees of freedom in the scalar-tensor gravity and
$f(R)$ gravity, the simplicity of the ansatz for the relation
between the entropy and horizon area is not enough to reproduce
corresponding Friedmann equations in those theories, in
particular, that ansatz is not enough to contain the dynamical
information of new degrees of freedom. However, if we move these
new degrees of freedom to matters in Einstein gravity, every
things then work very well. Therefore our result is useful to
further understand the holographic properties of non-Einstein
gravity theories. In addition, by using entropy formulas
(\ref{17}) and (\ref{34}), it should be of much interest to study
the generalized second law of thermodynamics of the FRW universe
in the scalar-tensor gravity and $f(R)$ gravity.

\section*{Acknowledgments}
We thank the referee for his/her useful comments which help us
improve the manuscript.  This work was supported in part by a
grant from Chinese Academy of Sciences, and grants from NSFC,
China (No. 10325525 and No. 90403029).



\end{document}